\documentclass{article}
\usepackage{ijcai20}

\usepackage[utf8]{inputenc}
\usepackage{amsthm}
\usepackage{amsmath}
\usepackage{amssymb}
\usepackage{graphicx}
\usepackage[small]{caption}
\usepackage{subcaption}
\usepackage{algorithm}
\usepackage{algorithmic}

\usepackage{times}
\usepackage{soul}
\usepackage{url}
\usepackage[hidelinks]{hyperref}
\usepackage{booktabs}
\usepackage{enumitem}

\usepackage{bbm}

\newtheorem{proposition}{Proposition}

\theoremstyle{definition}
\newtheorem{example}{Example}

\newcommand{\TN}{\textrm{TN}}
\newcommand{\TP}{\textrm{TP}}
\newcommand{\FN}{\textrm{FN}}
\newcommand{\FP}{\textrm{FP}}

\usepackage{color}
\usepackage[textsize=scriptsize]{todonotes}
\definecolor{blue}{RGB}{0, 93, 170}			%
\definecolor{darkgreen}{RGB}{0, 102, 0}

\newcommand{\citet}[1]{\citeauthor{#1}~\shortcite{#1}}
\newcommand{\citep}{\cite}

\usepackage{xspace}
\newcommand{\peer}{\text{\textsc{PeerNomination}}\xspace}

\title{\peer: Relaxing Exactness for Increased Accuracy in Peer Selection}

\author{
Nicholas Mattei$^1$
\And
Paolo Turrini$^2$\And
Stanislav Zhydkov$^{2}$
\affiliations
$^1$Department of Computer Science, Tulane University\\
$^2$Department of Computer Science, University of Warwick
\emails
nsmattei@tulane.edu,
\{p.turrini, s.zhydkov\}warwick.ac.uk
}

\hypersetup{draft}

\begin{document}

\maketitle

\begin{abstract}
In peer selection agents must choose a subset of themselves for an award or a prize. As agents are self-interested, we want to design algorithms that are impartial, so that an individual agent cannot affect their own chance of being selected. This problem has broad application in resource allocation and mechanism design and has received substantial attention in the artificial intelligence literature. Here, we present a novel algorithm for impartial peer selection, \peer, and provide a theoretical analysis of its accuracy.  Our algorithm possesses various desirable features. In particular, it does not require an explicit partitioning of the agents, as previous algorithms in the literature. We show empirically that it achieves higher accuracy than the exiting algorithms over several metrics.
\end{abstract}

\section{Introduction}\label{sec:introduction}

Peer selection, where agents must choose a subset of themselves for an award or a prize, is one of the pillars for quality assessment in scientific contexts and beyond. While current methods rely on expert panels, there is increasing attention to how to design trustworthy mechanisms that improve the accuracy and reliability of the outcome, keeping the procedure simple and cheap. The latter is particularly relevant in open online courses \cite{piech2013tuned}, where hiring professional graders is prohibitively expensive. Indeed, even IJCAI 2020 is implementing a portion of this system, requiring authors who submit papers to agree to be reviewers themselves.

The importance of having an ``objective" assessment in conference reviewing has been brought to light by the famous NIPS experiment \cite{langford_2015,shah2018design}: of all papers submitted to NIPS 2014, 10\% were reviewed twice by two independent committees which, astonishingly, agreed on less than half of the accepted papers in their pool. Whether the outcome was due to bias, incompetence or simply well-thought disagreement is still unclear. What is clear though is that the current solutions show undesirable properties. %

Methods for {\em impartial} peer selection, where self-interested individuals assess one another in such a way that none of them has an incentive to misrepresent their evaluation, have a long standing tradition in economics, e.g., \cite{Douceur2009,Holzman2013,Dollar}, which has in turn encouraged several groups in artificial intelligence and computer science more broadly to investigate these problems, e.g., \cite{KLMP15,AFPT11a,XZSS19,Aziz2019}.

The interest in such methods has culminated in a pilot scheme by the US National Science Foundation (NSF) \cite{naghizadeh2013incentives}, called for by \citet{MerrifieldSaari}, in which each principal investigator (PI) was asked to rank 7 proposals from other PIs. The rankings were then combined using the Borda score with the additional truth-telling incentive of receiving a bonus the closer one gets to the average of the other reviewers' marks. Though this method is not impartial, and leads to a Keynesian beauty contest \cite{Key36a}, the results were encouraging.

Research in artificial intelligence and economics has led to a number of proposals for algorithms choosing a set of $k$ agents from amongst themselves, commonly known as the peer selection problem.  We review some of the most prominent ones here to which we will compare our proposal. %

 \begin{description}[itemsep=0em]

 \item [Credible Subset \cite{KLMP15}.]  In Credible Subset (CS), reviewers assign scores to their allocated proposals and the potential manipulators, i.e., the reviewers that could be within the $k$ funded ones, are also selected to be funded, with a given probability. While the system is strategy-proof, it will yield an empty set of funded proposals in a number of cases \cite{Aziz2016}.

 \item[Dollar Raffle \cite{Dollar}.]   The Dollar Raffle method (DR) consists of reviewers distributing a score in the interval [0,1] to their reviews rather than independently allocating them as in (CS).  %

 \item[Exact Dollar Partition \cite{Aziz2019}] Reviewers are clustered at random and rank peers in different clusters.  Using a randomized rounding scheme based on the the shares computed with the method of \citet{Dollar}, the top proposals of each cluster are selected, depending on their clusters' importance. Dollar Partition is strategy-proof and has been shown to be the most accurate available method \cite{Aziz2019}.

\end{description}

We compare our algorithm against two more basic procedures: Vanilla, which selects the $k$ agents with the highest total Borda score based on the reviews received; and Partition, which, instead, divides the agents into a set of clusters and selects a predetermined number of them from each (typically $k$ divided by the number of clusters) as rated by the agents from the other clusters. Notice that, unlike Partition, Vanilla is not impartial but is commonly used as a baseline for comparison.

Relevant recent developments with a different focus use voting rules to aggregate ranks  (e.g., $k$-Partite  \cite{KKKKP18a}, including the  Committee \cite{KKKKP18a} and Divide-and-Rank \cite{XZSS19}) algorithms. Other methods are approval-based but only focus on single agent selection: Permutation~\citep{FeKl14a} and Slicing~\citep{BNV14a}.  Additional work in this area also focuses on assignment and calibration issues \cite{wang2019your,LianMNW18}.

\smallbreak
\noindent
\textbf{Our Contribution.} 
We present \peer, an impartial peer selection method for scenarios where $n$ agents review and are reviewed by $m$ others, with the goal of selecting $k$ of them. Each proposal is considered independently and it is selected only if it falls in the top $\frac{k}{n}m$ of the majority of its reviewers' (partial) rankings, using a probabilistic completion if such number is not an integer. 
This way we relax the exactness requirement, in the sense that our algorithm is not guaranteed to select exactly $k$ proposals every time.
However, under some mild rationality assumptions, the algorithm does so in expectation. Unlike other well-known peer reviewing methods, e.g., Exact Dollar Partition (EDP), \peer does not rely on clustering nor on reviewers submitting complete rankings, allowing more flexibility in where and when it may be deployed. %

We compare the performance of \peer against an underlying ground truth ranking when agent rankings are drawn according to a Mallows model \cite{Mal57,Xia19}, exactly deriving its expected accuracy analytically. Moreover, we empirically compare our method against other peer selection mechanisms, for which analytic performance bounds are unknown, using a number of well-known classification measures. Our results show that \peer improves on the current best performance in terms of accuracy known from the literature and relies on milder assumptions on the underlying reviewer graph. This suggests that relaxing the exactness requirement in peer selection outcomes can give us an improved performance with respect to the accuracy of the accepted set.

\smallbreak
\noindent
\textbf{Paper Structure.}
In Section \ref{sec:preliminaries} we set up the basic terminology and notation. Section \ref{sec:model} presents our algorithm and its theoretical properties. Section \ref{sec:experiments} compares its accuracy against the main existing alternatives, under various metrics. %

\section{Preliminaries}\label{sec:preliminaries}

We work with a set of agents $\mathcal{N} = \{ 1, 2, ..., n \}$ and an order over them, induced by their index, which represents the final ranking the agents would have, if they were to be assessed objectively. We refer to this order as the {\em ground truth}. Each agent is assigned $m$ other agents to review and is in turn reviewed by $m$ others.
We represent such $m$-regular assignment as a function $A: \mathcal{N} \rightarrow 2^{\mathcal{N}}$ and denote $i$'s review pool as $A(i)$, while $A^{-1}(i)$ denotes $i$'s reviewers. It is worth noting that while generating a random $m$-regular assignment is easy for small $m$ (by generating an $m$-regular bipartite graph),  sampling one uniformly is non-trivial and is an active area of study (e.g., see \cite{berger2010uniform}). In this paper, we assume uniform sampling to make our theoretical analysis tractable in Section \ref{sec:model} but not for experiments in Section \ref{sec:experiments}. In practice, we observed negligible effect on the performance of algorithms when using different assignment-generating procedures.
In real-world settings, agents can only review a limited number of proposals or papers so $m$ is typically small and constant, given $n$. 

Each reviewer $i$ submits a ranking of their review pool $A(i)$, which we represent as a strategy $\sigma_i: A(i) \rightarrow \{1, ..., m \}$, where $\sigma_i(j)$ gives the rank of $j$ given by $i$ in $i$'s review pool.
 A collection of all declared strategies
is called a \textit{profile} and is denoted by $\sigma$. The unique profile which is consistent with the ground truth is called \textit{truthful}.
After the individual preferences are declared, they are aggregated to select $k$ individuals.
We call a peer selection mechanism \textit{impartial} or \textit{strategyproof} if no agent can affect their chances of selection in any assignment using any strategy. 

\section{\peer}\label{sec:model}

In this section we present \peer and describe its performance analytically.

\subsection{The Algorithm}

A usual requirement for peer selection mechanisms is that it must return an accepting set exactly of size $k$ \cite{Aziz2019,AFPT11a,KKKKP18a}. Though some approaches investigated relaxing this assumption \cite{Aziz2016,KLMP15}, most notably the results by \citet{bjelde_fischer_klimm_2017} show that this relaxation can lead to better optimality approximation. We use this intuition in designing the following algorithm that returns an accepting set of size $k$ in expectation.

\peer works as follows: suppose every agent reviews and is reviewed by $m$ other agents. If an agent is in the true top $k$, we expect them to be ranked in the top $k$ proportion (i.e., top $\frac{k}{n} m$) of their review pool by the majority of agents that review them, if these were to report their accurate rankings. We say that an agent is \textit{nominated} by a reviewer if they are in the top $k$ proportion of the reviewer's declared ranking, i.e., their review pool. Likewise, we refer to $\frac{k}{n} m$ as the \textit{nomination quota}. Hence, for every agent $j$, we look at all reviewers $i_1, ..., i_m$ reviewing $j$ and select $j$ only if they are nominated by the majority of these reviewers. 

As $\frac{k}{n} m$ is unlikely to be an integer, we consider an agent {\em nominated for certain} if they are among the first $\lfloor \frac{k}{n} m \rfloor$ agents in the review pool, where $\lfloor x \rfloor$ denotes the whole part of a positive real number $x$. If they are in the next position (i.e., $\lfloor \frac{k}{n} m \rfloor + 1$), we randomly consider them nominated with probability $\frac{k}{n} m - \lfloor \frac{k}{n} m \rfloor$, that is, the decimal part of the nomination quota. Lastly, if the number of review pools an agent is part of is even, we require them to be nominated by just half of the review pools, not a strict majority. 

A crucial observation is that, since each agent is considered independently for selection, the algorithm is not guaranteed to return exactly $k$ agents. However, we will show that the algorithm is close enough to such number if the reviewers submit reviews that are close enough to the ground truth and, moreover, that truth-telling is an equilibrium outcome, i.e., \peer is impartial. 

The \peer algorithm is presented in Algorithm \ref{algorithm}. Note that in the algorithm we introduce the \textit{slack parameter} $\varepsilon$, which extends the nomination quota accordingly. As we show next, this is necessary in some settings to achieve the right expected size of the accepting set. 

\begin{algorithm}[h]
\begin{algorithmic}{}
\REQUIRE Assignment $A$, review profile $\sigma$, target quota $k$, slack parameter $\varepsilon$
\ENSURE Accepting set $S$
\STATE Set $\textit{nomQuota} := \frac{k}{n} m +\varepsilon$
\FORALL{$j$ in $\mathcal{N}$}
    \STATE Initialise \textit{nomCount} := 0
    \FORALL{$i \in A^{-1}(j)$}
        \IF{$\sigma_i(j) \leq \lfloor \textit{nomQuota} \rfloor$}
            \STATE increment \textit{nomCount} by 1
        \ELSIF{$\sigma_i(j) = \lfloor \textit{nomQuota} \rfloor + 1$}
            \STATE increment \textit{nomCount} by 1 with probability $\textit{nomQuota} - \lfloor \textit{nomQuota} \rfloor$
        \ENDIF
    \ENDFOR
    \IF{$\textit{nomCount} \geq \lceil \frac{m}{2} \rceil$}
        \STATE $S \leftarrow j$
    \ENDIF
\ENDFOR
\RETURN $S$
\end{algorithmic}
\caption{\peer}
\label{algorithm}
\end{algorithm}{}

\subsection{Expected Size and Slack Parameter}

We now derive the expected size of the accepting set returned by \peer as a function of $n, m$ and $k$. %
Since each agent is considered independently, we just need to derive the probability of selection for an agent given their ground truth position. Assume the algorithm is run on an $m$-regular assignment and the reviews are truthful. Note that we assume such assignment is sampled uniformly and so each review pool is equally likely to be assigned to any reviewer. Firstly, consider the probability of obtaining position $y$ in the sample of size $m$, given position $r$ in the underlying ranking. When drawing the sample, we need to choose $y-1$ individuals out of $r-1$ that are above agent $r$ in the ground truth, and then choose $m-y$ out of $n-r$ that are worse. In total, as expected, we are choosing $m-1$ other agents out of $n-1$. Hence:
\begin{equation*}
    \mathbb{P}[Y = y | R = r] = {r-1 \choose y-1} {n-r \choose m-y} \Big/ {n-1 \choose m-1}
\end{equation*} 
where $Y$ is a random variable representing the position in the review pool and $R$ is a random variable representing the ground truth position.
 
Denote now the nomination quota by $k_q := \frac{k}{n} m$ and recall that in any given review pool, top $\lfloor k_q \rfloor$ agents are nominated for certain and the next position is nominated with the probability of $k_q - \lfloor k_q \rfloor$. Hence, the probability of being nominated in any pool from position $r$ in the ranking is, independently:
\resizebox{.9\linewidth}{!}{
  \begin{minipage}{\linewidth}
 \begin{equation}
 \begin{split}
     q_r &:= \sum_{y=1}^{\lfloor k_q \rfloor} \mathbb{P}[Y = y | R = r] \\
     &+ \left(k_q - \lfloor k_q \rfloor \right) \mathbb{P}[Y = \lfloor k_q \rfloor + 1 | R = r]
 \end{split} 
 \label{eq:q_r}
 \end{equation}
 \end{minipage}
 }
\vspace{0.5em}

Since each review pool can be regarded as a Bernoulli trial with probability $q_r$ and to be accepted an agent has to be nominated $\lceil m/2 \rceil$ times, the probability of being accepted from position $r$ is given by the cumulative Binomial distribution:
\begin{equation}\label{eq:prob}
    \mathbb{P}[\textrm{accept} | R = r] = \sum_{i= \lceil m/2 \rceil}^m {m \choose i} q_r^i (1-q_r)^{m-i}
\end{equation}

\begin{figure}[h]
    \centering
    \includegraphics[width = .80\linewidth]{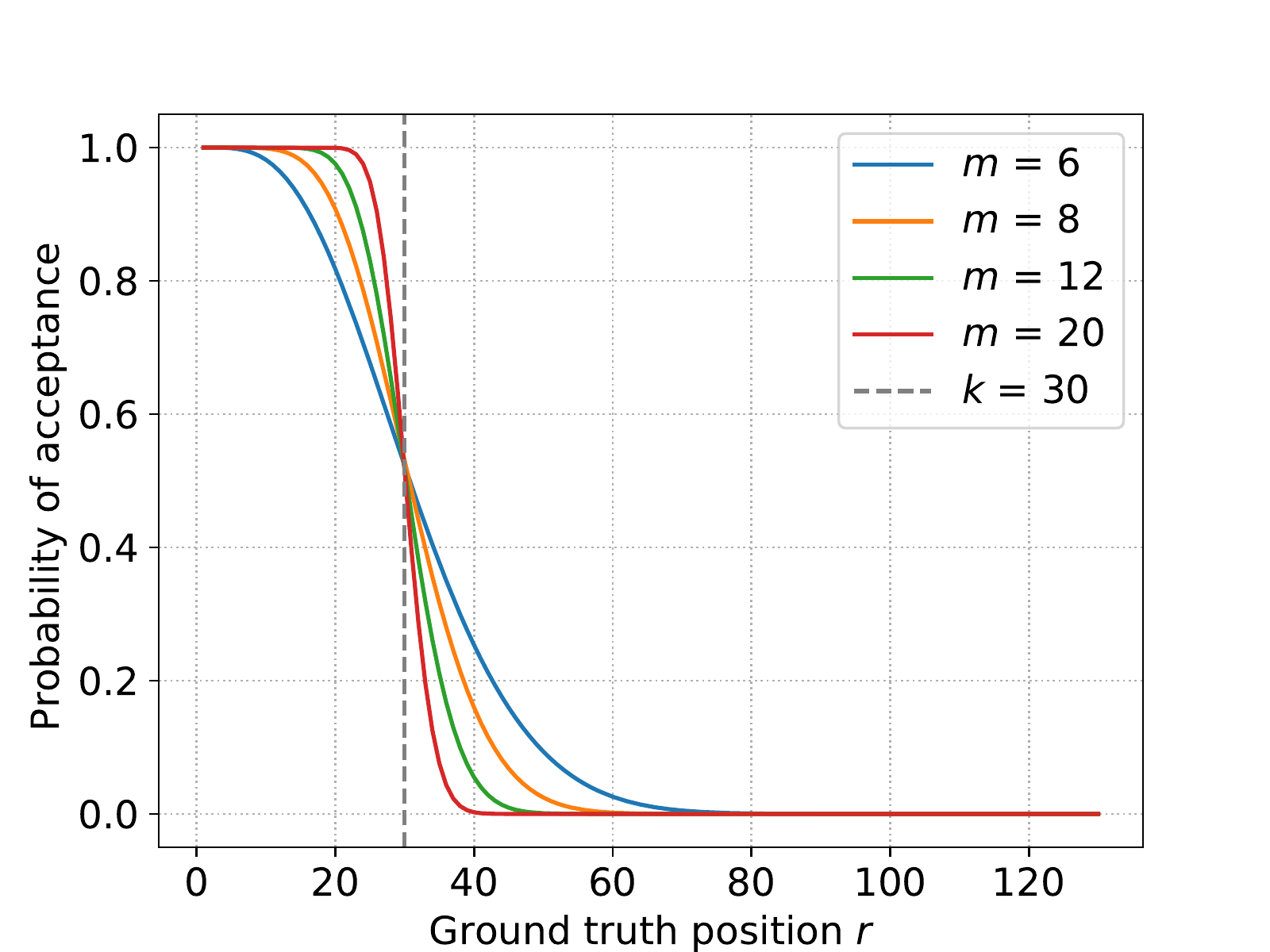}
    \caption{Probability of being accepted by the algorithm given the position in the ranking when  $n=130$ and $k=30$.}
    \label{fig:review_dist}
\end{figure}

An illustration of acceptance probabilities as a function of the ground truth position is shown in Figure \ref{fig:review_dist}. We can see that agents that are well inside top $k$ are almost certain to be accepted while those well outside of top $k$ are almost certain to be rejected. The width of the interval around top $k$ for which the probability is away from the extremes is dictated by $m$. Higher $m$ reduces uncertainty by providing more ``trials" for each agent and so narrows the interval.

We can now use the derived probability of acceptance to calculate the expected size of the accepting set. %

\begin{figure}
\centering
\begin{subfigure}[t]{0.48\linewidth}
	\centering
    	\includegraphics[width=0.98\linewidth]{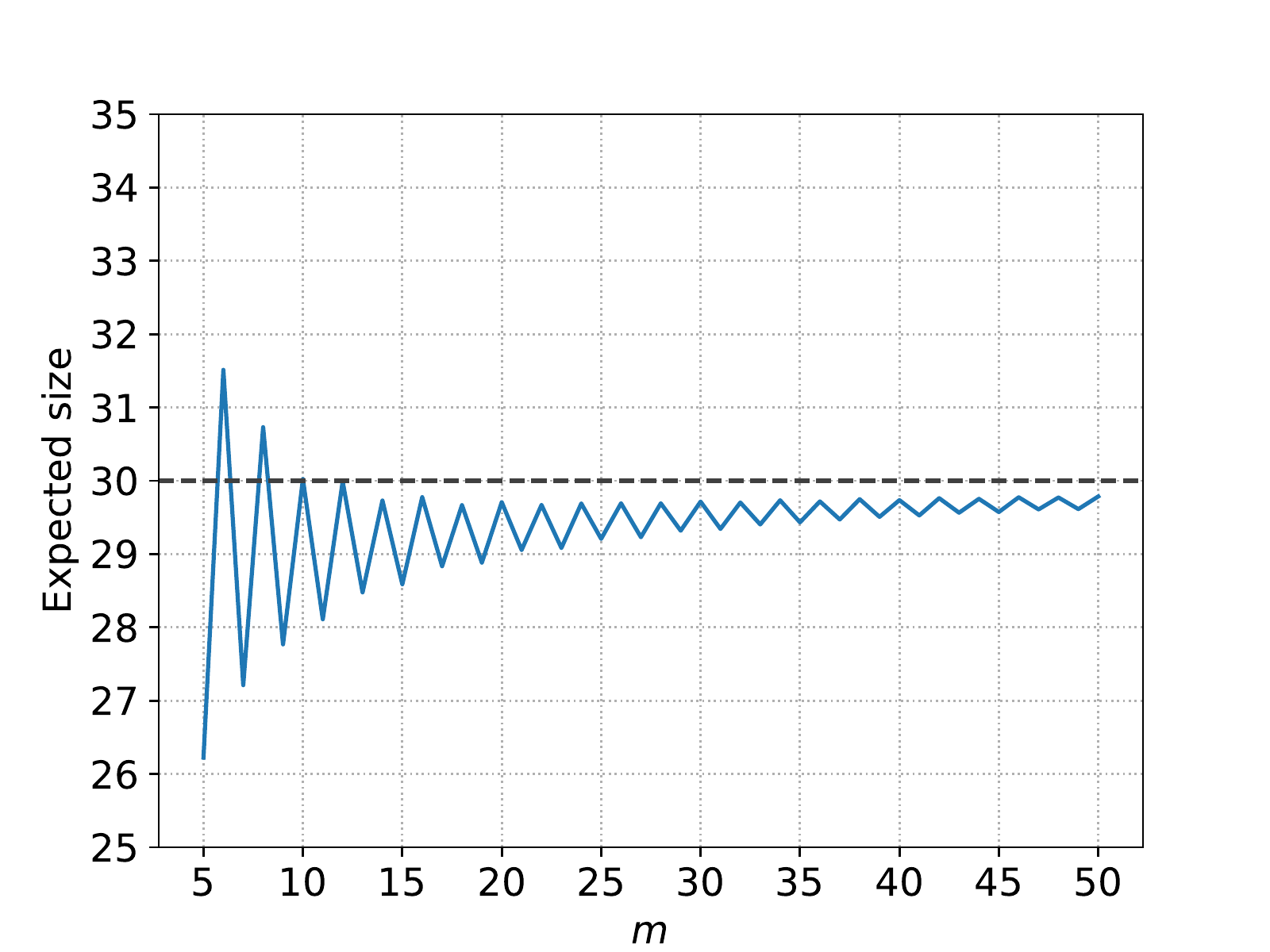}
    	\caption{Expected Size}
    	\label{fig:expected_size}
\end{subfigure}%
\begin{subfigure}[t]{0.48\linewidth}
	\centering
	\includegraphics[width = 0.98\linewidth]{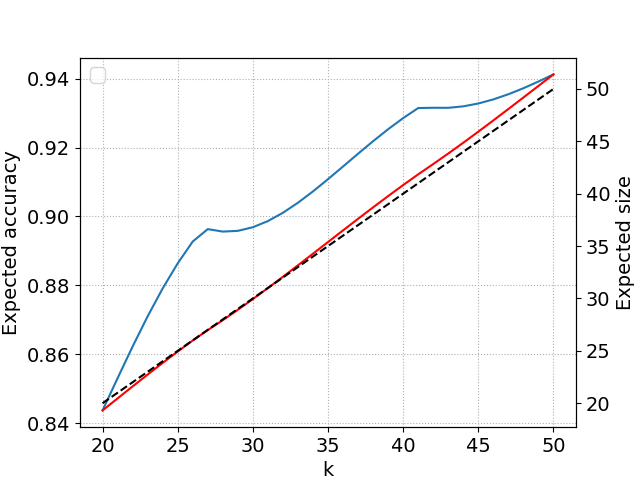}
	 \caption{Expected Accuracy}
    \label{fig:accept_exp_k}
\end{subfigure}	
\caption{(a) Expected size of the accepting set returned by the algorithm when $n=130, k=30$ and varying $m$.  (b) Expected accuracy and accepting size for different values of $k$. $n=130$, $m=9$ and $\varepsilon = 0.15$ were used for this figure. The red line shows the expected accepting size and the blue line shows the accuracy.}
\label{fig:size_accuracy}
\end{figure}

Since every individual is accepted independently with probability $\mathbb{P}[\textrm{accept} | R = r]$ and contributes 1 to the size if they are accepted, the expectation is simply $\sum_{r=1}^n \mathbb{P}[\textrm{accept} | R = r]$. The complexity of this expression makes it difficult to analyse it explicitly. However, Figure \ref{fig:expected_size} shows a typical behaviour of the expected size as a function of $m$. We observe that this approaches $k$ as $m$ increases. However, for small values of $m$ the expected size can vary significantly from $k$, especially when $m$ is odd (recall that agents need to get a clear majority in this case, making selection more difficult).
To tackle these issues, we introduce an additional parameter $\varepsilon$ that allows us to control the size of the accepting set more finely. If $\varepsilon$ is set to a non-zero value (usually a positive one), we extend the nomination quota in each review pool by this amount. Usually this increment simply contributes to the probability that the ``fractional nominee" is nominated. For example, in the setting $n=130, m=9$ and $k=30$, Figure \ref{fig:expected_size} shows the expected size slightly above 27 while our aim is 30. Setting $\varepsilon = 0.13$ yields the expected size very close to $30$. For most practical applications $\varepsilon \in [-0.05, 0.15]$, meaning the original algorithm is rather well-behaved.  Note that this is in contrast to other inexact mechanisms in the literature: Credible Subset must return no solutions with positive probability \cite{KLMP15}, while the Dollar Partition method may return as many additional agents as the number of clusters \cite{Aziz2016}.

Recall that the above analysis assumes reviewers to be accurate. If this assumption fails, we cannot provide any guarantees even for the expected size of the accepting set. It is also easy to construct marginal cases in which everyone or no one is selected in the worst case scenario.

\begin{example}
Consider the setting with 3 agents with everyone reviewing each other and suppose we want to select one individual (i.e., $n=3, m=2$ and $k=1$). Suppose agent 1 reviews 2 above 3, agent 2 reviews 3 above 1 and agent 3 reviews 1 above 2. The nomination quota with $\varepsilon = 0$ is $\frac{2}{3}$ and every agent is ranked in the first place once. Hence, each agent is selected with probability $\frac{2}{3}$ independently and so there exists a realisation where no one is selected as well as one where everyone is selected.
\end{example}{}
In Section \ref{sec:experiments} we consider a realistic setting that includes a noise model for the reviews and discuss the accepting size and the performance of \peer.

\subsection{Expected Size and Accuracy}
Above we derived the probability of acceptance given a position in the ground truth, assuming no noise, before introducing the parameter $\varepsilon$. It is easy to adapt this expression to include $\varepsilon$: simply update the nomination quota when computing $q_r$ in Equation \ref{eq:q_r}. Hence, let $k^\varepsilon_q = k_q + \varepsilon$ and

\resizebox{.9\linewidth}{!}{
  \begin{minipage}{\linewidth}
\begin{equation*}
\begin{split}
    q_r^\varepsilon &:= \sum_{y=1}^{\lfloor k^\varepsilon_q \rfloor} \mathbb{P}[Y = y | R = r] \\
    &+ \left(k^\varepsilon_q - \lfloor k^\varepsilon_q \rfloor \right) \mathbb{P}[Y = \lfloor k^\varepsilon_q +1 \rfloor | R = r]
\end{split}
\end{equation*}
\end{minipage}
}
\vspace{0.5em}

This gives us $\mathbb{P}[\varepsilon\textrm{-accept} | R = r]$ for each ground truth position by simply replacing $q_r$ in Equation \ref{eq:prob} by $q^\varepsilon_r$. The expected size is again given by a similar expression:

\begin{equation*}
    \mathbb{E}[\textrm{accepting size}] = \sum_{r = 1}^{n} \mathbb{P}[\varepsilon\textrm{-accept} | R = r]
\end{equation*}{}

It is now in principle easy to derive the expected accuracy of the algorithm. However, since the algorithm's output is inexact, there are multiple accuracy measures to consider, as is often the case for classification algorithms \cite{bishop2006pattern}. For example, we might care about how many agents of the true top $k$ we have selected (recall) or that we do not select too many agents from outside of it (false positive rate). We focus on the former, which we note is elsewhere referred to as {\em accuracy} \cite{Aziz2019}. The connection with classification metrics will be further explored in Section \ref{sec:experiments}. 
Now, the {\em expected recall} is simply the sum of the probability of selection over all true top $k$ positions, divided by $k$:

\begin{equation*}
    \mathbb{E}[\textrm{recall}] = \frac{1}{k} \sum_{r = 1}^{k} \mathbb{P}[\varepsilon\textrm{-accept} | R = r]
\end{equation*}{}

Again, the complexity of these expressions hinders theoretical analysis but Figure \ref{fig:accept_exp_k} shows a typical output for different values of $k$. 

While its performance appears good in isolation, it is important to compare \peer with other peer selection mechanisms which we do in Section \ref{sec:experiments}.

\subsection{Strategyproofness and Monotonicity}

Our main desired property is that of impartiality or strategyproofness. Luckily, this comes almost for free since the agents are chosen independently.

\begin{proposition}
The mechanism is strategyproof, i.e., no agent can affect their chances of selection using any strategy.
\end{proposition}{}

We also want the algorithm to be \textit{monotonic},  having better reviews does not hurt the chances of selection.

\begin{proposition}
The mechanism is monotonic, i.e., if a reviewer increases their ranking of an agent, that agent's probability of selection is not decreased.
\end{proposition}{}

\begin{proof}
Suppose $j$ is reviewed by $i$ and consider the probability of selecting $j$ given the original review of $i$, and a modified one where $j$ is ranked higher. There are three cases:
\begin{enumerate}[itemsep=0em]
    \item $j$ was already inside the integer part of the nomination quota in the original review or $j$ is still completely outside of the the nomination quota in the modified review. In both cases $j$ was already certain to be nominated or not nominated, respectively, by $i$, hence their probability does not change.
    \item $j$ moves from being a fractional nominee to being a full nominee increasing the chances of nomination (by $1 - (k_q - \lfloor k_q \rfloor)$), hence increasing their chances of selection.
    \item $j$ moves from being not nominated to be fractionally nominated increasing the chance of nomination (by $k_q - \lfloor k_q \rfloor$), hence increasing the chances of selection.
\end{enumerate}{}
In all cases $j$'s chances of selection do not decrease, completing the proof. %
\end{proof}{}

Notice that in the definition of the algorithm we stipulate that $\varepsilon$ is part of the input. One could be tempted to calculate $\varepsilon$ after collecting the reviews in order to adjust the output size to be exactly $k$, however this is undesirable for several reasons. Firstly, the run of the algorithm is non-deterministic, hence it might be impossible to find a value of $\varepsilon$ that guarantees such output size on every run. Secondly, and most importantly, this would eliminate strategyproofness since now an agent could estimate that reporting an untruthful review could decrease the size of the accepting set, hence forcing the mechanism to increase $\varepsilon$ and so increase their chances of selection.

\section{Simulation Experiments}\label{sec:experiments}

We draw a novel connection between inexact peer selection and the literature on classification in machine learning \cite{bishop2006pattern}.  With this empirical framework we run experiments to demonstrate that \peer outperforms other mechanisms proposed.

\subsection{Classification Measures}

The usual and intuitive way to measure the ``accuracy'' of an exact peer-selection mechanism is counting how many agents from the top $k$ positions in the ground truth have been selected, as a proportion of all $k$ agents selected. This allows us to compare exact peer-selection mechanisms as was done in \cite{Aziz2019}. However, comparison with inexact mechanisms is less obvious. Since the accepting set is not guaranteed to be of size $k$ exactly, any output with more than $k$ agents may artificially increase the accuracy of the inexact mechanism and the opposite for any smaller output. One option is to measure the accuracy as a proportion of the output size, however, this approach will overrate outputs that are accurate but much smaller than $k$. 

Inexactness allows us to view peer selection as a classification problem in which selection means positive classification. We can then view the selected agents from the true top $k$ as true positives and the non-selected agents from outside the true top $k$ as true negatives. We apply the standard classification accuracy measures \cite{bishop2006pattern} such as recall and precision to \peer to analyse its performance. 

More formally, let $S$ be the set of agents selected by the algorithm and $S^+ = \{r \in S \mid \textrm{rank}(r) \leq k \}$ the set of selected agents that are in the true top $k$, i.e., true positives (TP). Similarly, we can use $S^- = \{r \in S \mid \textrm{rank}(r) > k \}$ for false positives (FP).
Hence we can define:
    $\textrm{TP} = |S^+|$,
    $\textrm{FP} = |S^-| = |S| - \textrm{TP}$, 
    true negatives $\textrm{TN} = | \{ r \notin S \mid \textrm{rank}(r) > k \} | = n - k - \FP$, and
    false negatives $\textrm{FN} = | \{ r \notin S \mid \textrm{rank}(r) \leq k \} | = n - |S| - \textrm{TN}$.

We can now look at some of the normal performance metrics: {Positive Predictive Value (PPV) (aka Precision)}, {True Positive Rate (TPR) (aka Recall)} and  \textrm{False Positive Rate (FPR)}, defined as follows: %
\noindent

\resizebox{.9\linewidth}{!}{
  \begin{minipage}{\linewidth}
\begin{align*}
    & \textrm{PPV} := \frac{\TP}{\TP + \FP} & &
    \textrm{TPR} := \frac{\TP}{\TP + \FN} & &
 \textrm{FPR}  := \frac{\FP}{\TN + \FP}
\end{align*}
\end{minipage}
}
\vspace{0.5em}

Furthermore, we can view the slack parameter $\varepsilon$ as the sensitivity threshold akin to the probability threshold in the machine learning literature (see e.g., \cite{flach}). %

\begin{figure}[!h]
\centering
    \begin{subfigure}[t]{0.48\linewidth}
        \centering
        \includegraphics[width=.98\linewidth]{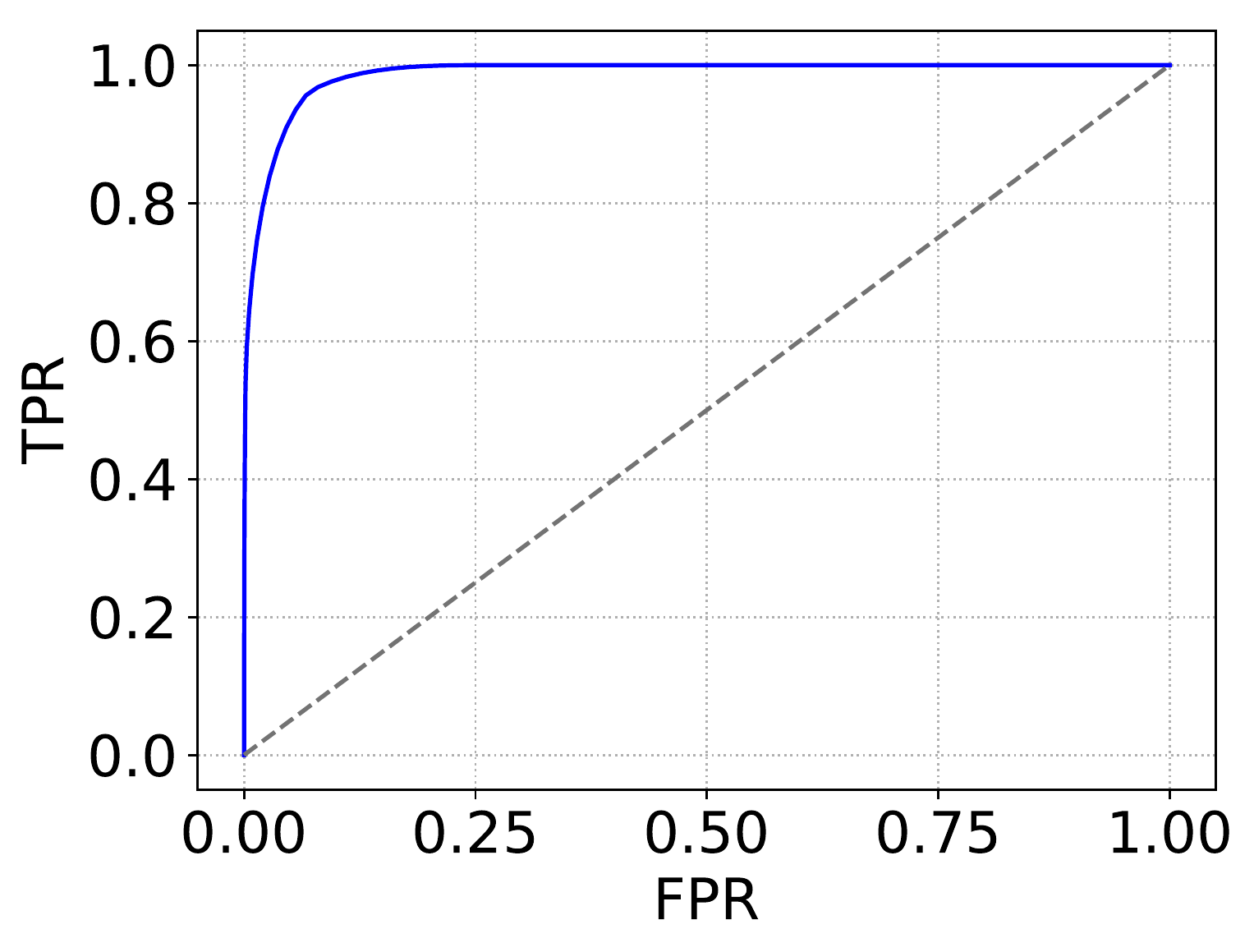}
        \caption{ROC Curve}
\end{subfigure}%
\begin{subfigure}[t]{0.48\linewidth}
        \centering
        \includegraphics[width=0.98\linewidth]{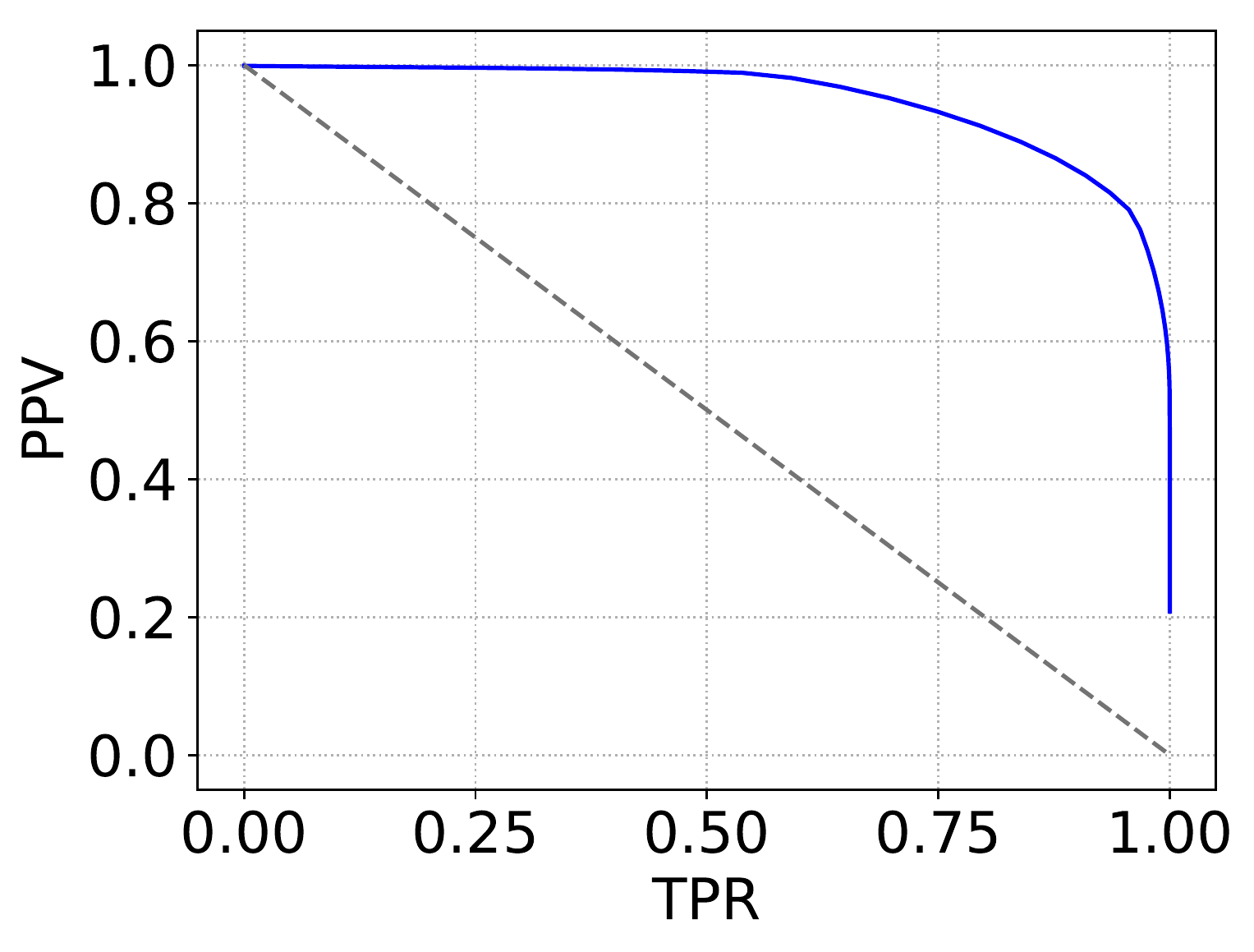}
        \caption{PR Curve \\\hspace{\textwidth}}
    \end{subfigure}
    \vspace{-1.5em}
    \caption{ROC and PR curves for \peer. They were computed analytically with $n=120, m=8, k=25$.}
    \label{fig:roc}
\end{figure}

This suggests a method to construct the Precision-Recall (PR) and Receiver-Operator Characteristic Curve (ROC): vary $\varepsilon$ such that the nomination quota varies between 0 and $m$ and measure the Precision, Recall and False Positive Rate at each value. An example is presented in Figure \ref{fig:roc}.

The curves show the trade off between sensitivity (TRP) and inclusivity (FPR). %
As we follow the ROC curve, which corresponds to gradually increasing the nomination quota, the (TPR) increases quickly, i.e., we do not need to accept \textit{too many} extra agents to select all the deserving agents. On the other hand, we can still achieve TPR of around 0.8 with the FPR very close to 0. This shows that we can select around 80\% of the agents in the true top $k$ if we concentrate on not selecting the ``undeserving" individuals. While the curves are interesting on their own, we want to be able to compare them to other peer-selection mechanisms, so an important direction is finding a generalizable way of constructing curves for other peer-selection mechanisms.

\begin{figure*}
    \centering
    \includegraphics[width=.85\linewidth]{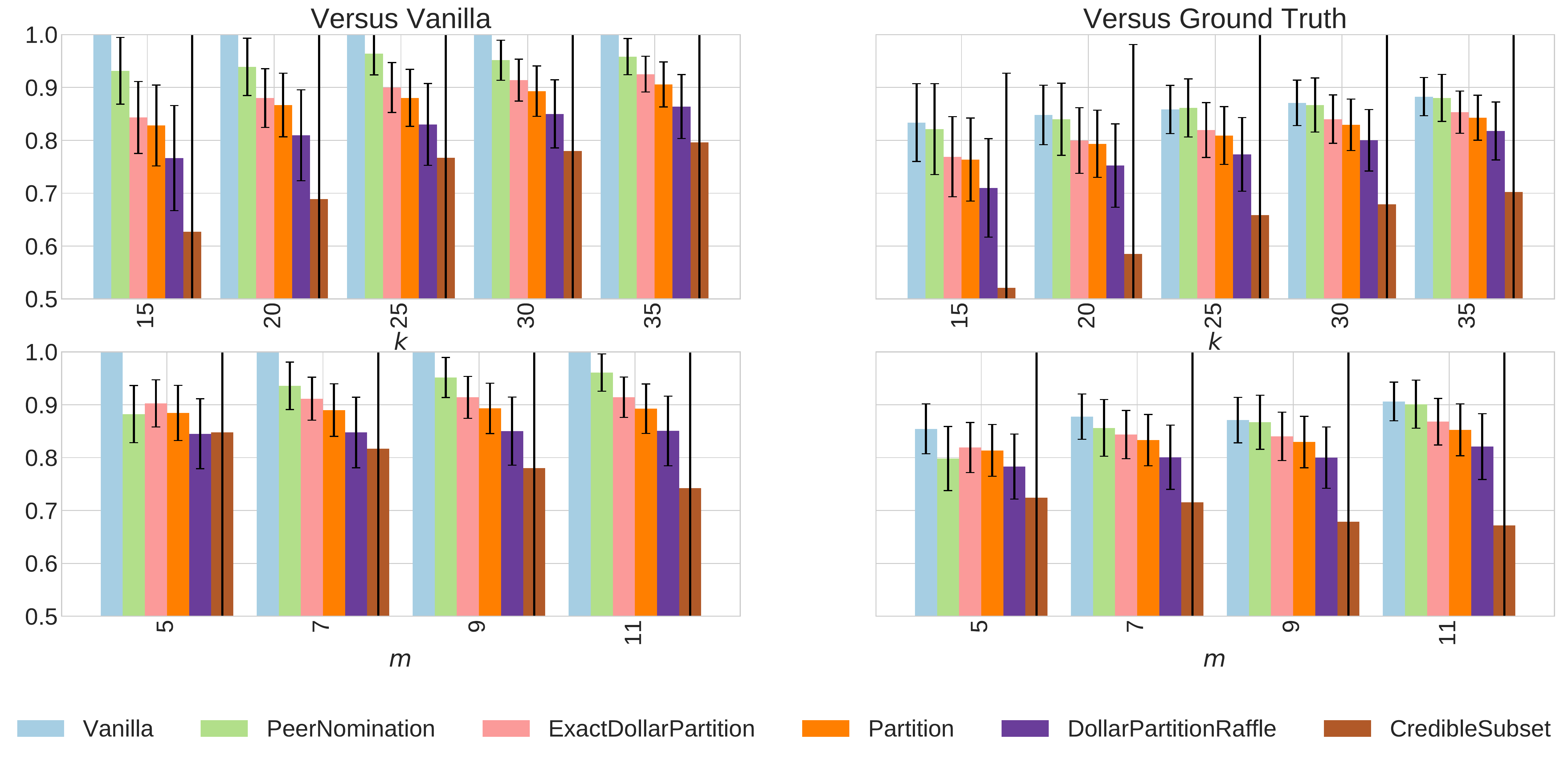}
    \caption{Comparison of prominent algorithms against a Vanilla baseline (left) and against the ground truth ranking of a Mallows Model (right). $n=120, l=4, \varphi=0.5$. On top $m=9$. On the bottom $k=30$.  \peer out performs across settings except $m=5$.}
    \label{fig:results}
\end{figure*}{}

\subsection{Experimental Setup}

We extend the testing framework developed by \citet{Aziz2019} and using methods from \textsc{PrefLib} \cite{MaWa17,MaWa13a}.  Our code and data is available online \footnote{\url{https://github.com/nmattei/peerselection}}. As in \citet{Aziz2019}, we set $n=120$ and tested the algorithm on various values of $k$ and $m$. The test values for $k$ were $15, 20, 25, 30, 35$ and the test values for $m$ were $5,7, 9, 11$. For the algorithms that rely on the partition, we chose the number of partitions, $l$, to be 4.

For each setting of the parameters we generated a random $m$-regular assignment matching reviewers to reviewees.  As in other works, we model he reviews of each agent using a Mallows Model \cite{Mal57}. In a Mallows model we provide a (random) ground truth ranking $\pi$ and a noise parameter $\phi$.  If we set $\phi=0$ then agents will always report $\pi$ as their ranking, i.e., they are all exactly correct.  As we increase $\phi$ agents will report increasingly inaccurate rankings as a function of the Kendall tau distance between $\pi$ and all possible rankings.  Note that each agent draws from this distribution independently.  Hence, by varying $\phi$ we can test the robustness of our algorithms to errors in the rankings submitted by the agents.  Mallows models have a long history in machine learning and group decision-making as they can simulate noisy observations of a ground truth ranking, and be sampled efficiently \cite{Mal57,Xia19,lu2011learning}.

\begin{figure}
    \centering
    \includegraphics[width=0.8\linewidth]{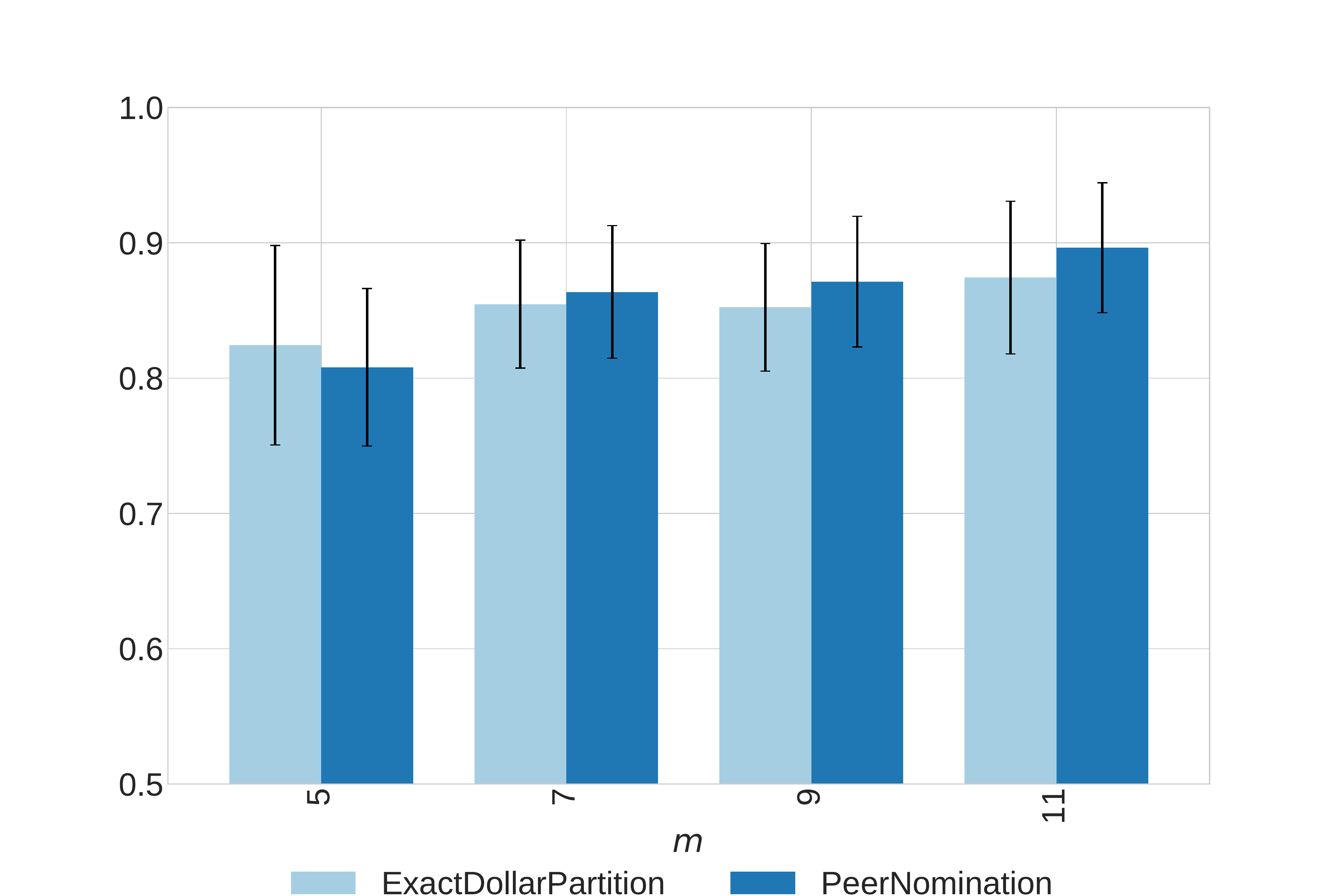}
    \caption{Results of the forced size experiment: \peer and EDP are always guaranteed to return the same number of agents.}
    \label{fig:fair_results}
\end{figure}

The experiment was repeated 1000 times for each setting, after which the average recall was calculated giving us high confidence in our results. For \peer, we used theoretical estimates of $\varepsilon$ to achieve the right expected size of the accepting set. 
The error bars in Figures \ref{fig:results} and \ref{fig:fair_results} represent 1 standard deviation of the data. In line with the observation in \cite{Aziz2019}, varying the dispersion parameter $\phi$ for Mallows' noise did not have significant effect on the accuracy of all algorithms until we approach $\phi = 1.0$ when all reviewers report a completely random ordering. The results for $\phi = 0.5$ are presented in Figure \ref{fig:results}.

\peer does not require an explicit partitioning making it more flexible.  Another issue with partitioning, as pointed out in \cite{Aziz2019}, is that the performance of both EDP and Partition degrade as we increase the number of clusters.  In another test we varied the number of clusters $\ell$ between 2 and 10.  We saw a decrease in performance of about $3-4\%$ for the partition based methods while the performance of \peer remained constant.

In another testing setup we adopted a slightly different  procedure in order to ensure a level comparison. In each simulation, we generate a random $m$-regular assignment, run \peer using the target $k$ as an input, measure the size of the output and run EDP using this size as the input $k$.  A similar experiment was also performed for the inexact version of Dollar Partition in \citet{Aziz2016}.  This ensures that during each simulation both algorithms return the same number of agents for selection. The results of this comparison are presented in Figure \ref{fig:fair_results}.

\subsection{Results}

Again following \citet{Aziz2019} and depicted in Figure \ref{fig:results}, we compared a selection of impartial peer selection algorithms and Vanilla (Borda count). Borda is a classic social choice rule that is known not to be strategyproof but is optimal in the ordinal peer-ranking setting under the assumption of no noise \cite{caragiannis_krimpas_voudouris_2016}, and thus represents an optimistic baseline in the presence of no manipulators. \peer outperforms EDP significantly in the the majority of the settings we have considered. The only setting where EDP outperforms \peer is at $m=5$, which is a low information setting, where reviewers are given a nomination quota fractionally above 1. However, our algorithm improves quickly with $m$. Even at $m=9$ shown in Figure \ref{fig:results} \peer approaches the performance of Borda across the values of $k$ we considered.

It is worth noting that \peer tends to return a slightly larger than $k$ set on average (usually $<1$ additional agent). Nevertheless, even if the testing forces \peer and EDP to return the same number of agents every time, we see that \peer has an overall advantage as shown in Figure \ref{fig:fair_results}. Again, EDP only does better in a low information setting ($m=5$).

\section{Conclusion}\label{sec:conclusion}

There are many avenues for future work: \peer, which already does not rely on predefined clustering, can be extended to not require an $m$-regular assignment. Moreover, each reviewer does not even need to declare a full ranking over their review pool, but simply declare the nominees for the selection and one nominee to be fractionally selected. This also suggests that there might be a possible extension of the algorithm which makes use of the declared rankings in full, as this data is currently discarded.

Crucially, the usefulness of our algorithm depends on returning an accepting set of size close to $k$. We saw that this can be achieved using the parameter $\varepsilon$. However, we saw that very high levels of noise can affect the size of the accepting set. This suggests that an important research direction will be that of testing different models of agent behaviour in detail.

{\small
\bibliography{ref}
\bibliographystyle{named}
}

\end{document}